# Mutiferroicity in a spin-chain compound, $Tb_2BaCoO_5$, with exceptionally large magnetodielectric coupling in polycrystalline form

**Sanjay K Upadhyay and E.V. Sampathkumaran**

*Tata Institute of Fundamental Research, Homi Bhabha Road, Colaba, Mumbai 400005, India*

**Abstract**

We report the results of detailed investigations of magnetization, heat-capacity, dielectric, pyrocurrent, and magneto(di)electric measurements on $Tb_2BaCoO_5$, belonging to a hither-to-unexplored spin-chain cobaltate family, $R_2BaCoO_5$ (R= Rare-earths). The magnetic measurements reveal that this compound exhibits an antiferromagnetic transition at ($T_N$=) 18.8 K and there is a spin reorientation beyond 40 kOe below $T_N$. Dielectric and pyrocurrent data measured as a function of temperature and magnetic-field establish that this compound is a 'type-II' multiferroic material. The most fascinating finding which we would like to emphasize is that the observed value of magneto-dielectric effect beyond the metamagnetic transition field is the largest (close to 55%, below 10 K, for $H$= ~100 kOe) ever reported for polycrystals of a compound in bulk form, thereby offering a hope to find a single-phase polycrystalline compound at room temperature to enable ease of applications.

In the research area exploring the control of ferroic order by another kind of ferroic order, historically in the 'so-called' type I' multiferroic materials, it has not been easy to find a single-phase polycrystalline compound with large magnetoelectric effect to enable applications with ease of synthesis. This is due to the weakness of the coupling between magnetic and electric dipoles[1] in these materials. Though this area of research got boosted when spin-driven ferroelectricity ('type II') was found in magnetic materials[1-7], single phase materials in polycrystalline form have been found to show small values of magnetodielectric (MDE) effect. In this respect, the best values reported till now are: 10% for $TbMnO_3$ (ref. [2]), 7% for $EuTiO_3$ (ref. [8]), 13% for $Ba_{0.5}Sr_{1.5}Zn_2Fe_{12}O_{22}$ (ref. [9]), 16% for $CaBaCo_4O_7$ (ref. [10]), and most recently 18% for $Tb_2BaNiO_5$ (ref. [11]). We report here that the compound, $Tb_2BaCoO_5$, established to exhibit 'type II' multiferroicity below 18.8 K in this work, exhibits significantly larger MDE coupling (about 55% below 10 K in ~100 kOe). Thus, this work brings out the existence of a strongly coupled multiferroic compound in the polycrystalline form in a hitherto unexplored family of cobaltates, offering a hope to find such multiferroics at room temperature.

It is now established that antisymmetric exchange interaction ($S_i$ x $S_j$) interaction, usually referred to as Dzyaloshinski-Moriya (DM) exchange interaction resulting in the breaking of inversion symmetry at low temperatures due to the spiral magnetic structure, exchangestriction mechanism, as proposed, for instance, for in $Ca_3CoMnO_6$ (ref. [12]), $Gd(Dy)FeO_3$ (ref. [13,14]) and/or spin-direction dependent metal-ligand hybridization are commonly responsible for triggering ferroelectricity. There is a growing realization that, even in those compounds which are centrosymmetric, loss of inversion symmetry necessary for ferroelectricity can be obtained by a certain behavior of magnetic ions (charge disproportionation or off-centering)[15,16] and local effects such as tilting of coordination polyhedra[17]. In this context, recent evidence for multiferroicity[11, 18-24] in centrosymmetric (*Immm*, orthorhombic) spin-chain compounds, $R_2BaNiO_5$ (R= Rare-earths), the well-known prototype series for Haldane's predictions, gains importance. Density-functional theory on the Er analogue establishes[23] the existence of a displacive component that distorts oxygen octahedra around Ni. Initially, the present investigations were intended to address a question whether the observation of multiferroicity in this family crucially depends on the Haldane spin-chain nature (that is, in $S$= integer systems). We present here a detailed investigation of a member (viz., $Tb_2BaCoO_5$) in the isostructural Co series, $R_2BaCoO_5$ (ref. [25-27]) by measurements of magnetization ($M$), heat-capacity ($C$), dielectric permittivity, bias current ($I_{bias}$) and pyrocurrent ($I_{pyro}$) as a function of temperature ($T$= 2-300 K) and magnetic field ($H$). It may be added that a preliminary study of magnetic susceptibility ($\chi$) behavior was carried out long ago[25].

The orthorhombic compounds with the formula, $R_2BaMO_5$ (M= Co, Ni, and Cu), have been known to exhibit dimorphism, depending on the $R$ ion, in *Immm* and *Pnma* space groups. The *Immm* space group, which is of interest to this article, is characterized by vertex-sharing $M$ chains (running along *a*-axis). These $M$ chains are isolated from each other by intervening $R$ and Ba ions (Fig. 1a). The $NiO_6$ octahedran in this space group is distorted, with the apical distance (1.88 Å) being shorter than the basal Ni-O distance (2.18 Å). Three dimensional ordering occurs, if $R$ is a magnetic-moment ion. In Co family, the *Immm* space group was found for $R$= Pr-Tm, and polymorphism was observed for $R$= Tb, Dy, Ho, Er or Tm.

Polycrystalline $Tb_2BaCoO_5$ was prepared by solid state reaction, starting with stoichiometric amounts of $Tb_2(CO_3)_3.nH_2O$ (purity >99.9%), $CoCO_3.nH_2O$ (>99.999%) and $BaCO_3$ (>99.9%). These oxides were thoroughly mixed together and heated in an alumina crucible in an atmosphere of argon at 1200 $^0$C for 20 h and at 1350 $^0$C for 30 h with an intermediate grinding. The sample was characterized by x-ray diffraction (Cu $K_\alpha$), which revealed that the specimen is single phase (in *Immm* space group) within the detection limit of the technique (<2%). A Rietveld refinement (Fig. 1b) of the pattern (using the FULLPROF program) yielded lattice constants [*a*= 3.756(2) Å, *b*= 5.825(1) Å, and *c*= 11.557(1) Å], agreeing with ref. [25]. For details of dc magnetization, *C(T)*, dielectric permittivity, $I_{bias}$ and $I_{pyro}$ measurements, the readers may see refs. [18-23].

The χ obtained in a field of 5 kOe (Fig. 2a) and 100 Oe (Fig. 2b, inset) exhibits a monotonous increase with decreasing temperature till about 20 K, which is cut off by a peak at 18.8 K, followed by a steep drop. This signals that there is an antiferromagnetic transition at ($T_N$=) 18.8 K. There is a weak upturn below about 5 K, which could be intrinsic to this compound or due to a trace of a magnetic impurity. The low-field curves obtained for zero-field-cooled (ZFC) and field-cooled (FC) conditions overlap, thereby ruling out spin-glass freezing. Inverse χ plot is linear above 50 K (Fig. 2) with a very weak deviation from this linearity at lower temperatures, which may be due to crystal-field effects and/or due to short-range magnetic order. The effective moment obtained from the high temperature linear region is 9.72 ± 0.05 $\mu_B$/Tb and this value corresponds to that of trivalent Tb ion. It therefore appears that the contribution from divalent Co to paramagnetic χ is negligible, despite the fact that Co in an octahedral environment should possess a spin of ½. Strong covalent mixing of Co 3d and (apical) O 2p orbitals, as reflected in the shorter bond distance of 1.88 Å, has been proposed as a cause of this negligible contribution[25] to χ, as noted for $R_2BaNiO_5$ family also[28]. The value of paramagnetic Curie-Weiss temperature (~ -20 K) is very close to the observed magnetic ordering temperature, with the negative sign being consistent with Néel order. In isothermal magnetization curves below $T_N$ (Fig. 2b), following linear variation at low-fields, there is a sudden upturn with increasing *H*, indicating the existence of a spin-reorientation (or metamagnetic) transition. This behavior of *M(H)* is consistent with the conclusion that the zero-field state below $T_N$ is antiferromagnetic. The field at which this upturn occurs is about 40 kOe for 2 K, gradually decreasing with increasing temperature. The curves around this field get gradually broadened with increasing *T*. These *M(H)* curves are found to be weakly hysteretic due to metamagnetic transition. The *M(H)* curves below $T_N$ are *S*-shaped, with the values at the highest field (140 kOe) being much smaller than that expected for fully degenerate trivalent Tb ion due to crystal-field effects. The values do not saturate even at 140 kOe. This *S*-shaped curve may also suggest that the spin realignment continues to be gradual in such higher fields.

In order to render further support to antiferromagnetic ordering, we show the results of heat-capacity measurements in figure 2c at low temperatures only (as the data at higher temperatures was found to be featureless). There is a clear λ-anomaly at $T_N$ (see zero-field data). The data below $T_N$ can be fitted to a $T^3$ functional form, which is consistent with antiferromagnetic spin-wave theories. The peak in *C(T)* shifts to a lower temperature with increasing fields (Fig. 2c). If *C(T)* is obtained in a field which causes spin-reorientation (say, in 60 kOe), the low-*T* data deviates from $T^3$-dependence, as expected. The zero-field and 60 kOe- curves merge above 30 K with reduced values for the latter for 30 K > *T* > $T_N$ and the isothermal magnetic entropy derived from these data is found to get reduced (by about 2J/mol $K^2$) above $T_N$. This finding confirms the existence of short-range magnetic order over a decade of temperature above $T_N$ in zero field.

We have measured dielectric permittivity with various frequencies in the range 5-100 kHz, as well as in the presence of external magnetic fields, with an ac bias of 1 V. The value of the loss term, tan$\delta$, is insignificant (<0.006) typical of very good insulators in the *T*-range of interest. Hence, any interference of extrinsic contributions to dielectric permittivity is ruled out, at least below 100 K. In dielectric constant (ε') as a function of $T$, there is no worthwhile feature above 20 K (Fig. 3a, inset), but there is a distinct peak at ~18.8 K, suggesting possible ferroelectric order at this temperature, coupled with magnetic order. Incidentally, the peak shifts to a marginally lower temperature (by 0.4 K) as the frequency is increased from 5 to 100 kHz, the origin of which is not clear. Further evidence for the coupling between magnetic and electric dipole is obtained from figure 3b, in which ε'($T$) is plotted in the presence of external magnetic fields. There is a significant shift of the peak temperature to a lower value with increasing $H$, with profound changes in absolute values of ε' below $T_N$, but with relatively small variations above $T_N$. In order to further substantiate this finding, it is instructive to show the magnetodielectric value Δε', defined as {ε'(H)-ε'(0)}/ε'(0), in figure 3c. There is a qualitative change in the shapes of the curves as the temperature is varied across $T_N$, with the values of Δε' remaining relatively low (close to a few percent) above $T_N$. For $T<T_N$, the value of Δε', for initial applications of magnetic field, varies very weakly. Beyond certain field, there is a dramatic increase in the value. The field at which this sharp upturn occurs decreases gradually with increasing $T$, and it essentially tracks the spin-reorientation field inferred from $M(H)$ data. Thus, there is an interdependent metamagnetic and metaelectric phenomenon occurring in this material. Therefore this compound can be classified as a 'metamagnetoelectric' material. There is a peak at a high field (e.g., at 100, 85, and 75 kOe at 7, 10, and 15 K respectively) in the plots below $T_N$, followed by a gradual decrease at further higher fields. Such a feature is absent above $T_N$, e.g., at 40 K. These features below $T_N$ – initial weak field dependence, followed by a sharp upturn, a peak and reversal in magnitude – mimic those seen for 10 atomic percent Yb doped $GdMnO_3$ (ref. [29]). In this case, it was proposed that both symmetric exchange interaction (resulting in exchangestriction) and asymmetric DM interaction may be operative. While at low fields the former mechanism (and/or possible local distortions due to magnetostriction) may be operative, the latter possibly gradually dominates[30] following spin-flop by the application of magnetic fields resulting in an enhanced magnetoelectric coupling at high fields. In addition, we believe that, with the application of very high fields beyond the peak fields, gradual suppression of DM interaction (that is, owing to the tendency towards ferromagnetic alignment at extremely high fields) results in the reversal in the values of Δε' (Fig. 3c). We are handicapped by the absence of detailed neutron diffraction studies to draw firm conclusions with respect to magnetic structure. Nevertheless, the most important point to be noted is that the peak values of Δε' reach as much as ~55% (say, at 7 K near 100 kOe), which is about 300% larger compared to that observed for the Ni analogue[11]. Though much larger values have been reported in single crystalline and thin film forms of different materials in the past literature, the ones for polycrystalline forms of magnetoelectric compounds hitherto studied in the literature never exceeded 20%, as mentioned in the introduction. Competing magnetic exchange interactions strongly influenced by magnetic field and anisotropy of the Tb 4f orbitals, may be the source of exceptional magnetodielectric behavior of the present compound.

In order to endorse that the dielectric peak at $T_N$ is due to ferroelectricity, we have performed $I_{bias}$ as well as $I_{pyro}$ measurements. Conventional $I_{pyro}$ measurement is done by cooling the specimen in an electric field, $E$, (to align the electric dipoles) and, after switching off this electric field at desired temperature, the capacitor is shorted for sufficient time to remove stray charges (if any); the $I_{pyro}$ is then measured as a function of $T$ while warming. It is now realized that inferences from such $I_{pyro}$ measurements could be sometimes misleading, as 'thermally stimulated depolarization current (TSDC)' can also yield $I_{pyro}$ peaks[31,32]. Recently, it was proposed[33] that $I_{bias}$ data is free from such an ambiguity; this involves cooling the specimen to a desired temperature in the absence of any electric field and then measuring the current as a function of temperature in the presence of an electric field. The influence of rate of variation of temperature can also distinguish between these two possibilities, as a feature due to TSDC would show a shift for different rates. We show the results of these measurements for an electric field of 3.5 kV/cm in figure 4. Both $I_{pyro}$ and $I_{bias}$ exhibit an asymmetric feature at $T_N$ typical of ferroelectricity and the peak temperature is insensitive to the rate of change of temperature (Fig. 4a). The plot gets inverted with a change in the polarity of electric field (Fig. 4a, inset) expected for ferroelectricity. All these findings establish that this compound is a type-II multiferroic material. We have also studied the behavior of $I_{bias}$ in the presence of external magnetic fields, and the remnant polarization curves derived from $I_{bias}$ curves are shown in figure 4b. As in the case of in-field $\varepsilon'(T)$ peak (Fig. 3b), the curve in general shifts towards lower temperatures, if the value of $H$ is increased. Onset temperature of polarization is depressed significantly after the metamagnetic transition field is crossed. The values well below $T_N$, say, at 5 K, are also suppressed by $H$ remarkably. These findings establish strong magnetoelectric coupling in this compound.

To conclude, we establish that the spin-chain compound, $Tb_2BaCoO_5$, is a 'type II' multiferroic compound below 18.8 K. The most notable observation is that the observed change in the dielectric constant below $T_N$ brought out by externally applied magnetic fields, following spin-reorientation, is the largest ever observed for bulk form of polycrystals of a compound. This work therefore offers a hope to find such polycrystalline single-phase materials with large MDE at room temperature. Ease of synthesis of bulk quantities of such polycrystals compared to thin films and single crystals may facilitate device applications. The results presented in this article point to rich physics behind this hitherto ignored family of spin-chain cobaltates, and should attract further research, just as some other cobaltates, e.g., $RCoO_3$, have been of interest to condensed matter community for the past several decades. Finally, considering that many members of the isostructural Ni-based Haldane spin-chain family, $R_2BaNiO_5$, have been reported to show multiferroic behavior, the present finding on the Co system establishes that integer-spin chain is not a necessary criterion to observe multiferroicity in this structure.

Figure 1:
(a) *Immm* orthorhombic crystal structure and (b) x-ray diffraction pattern of $Tb_2BaCoO_5$. Rietveld refinement, along with fitted parameters, is also shown.

Figure 2:
(a) Magnetic susceptibility ($\chi$) and inverse $\chi$ as a function of temperature, measured in a field of 5 kOe. The continuous line through the data points in the high temperature range in $\chi^{-1}$ versus $T$ plot is obtained by Curie-Weiss fitting. In the inset of (b), the low temperature behavior measured in 100 Oe for zero-field-cooled and field-cooled conditions of the specimens is shown. (b) Isothermal magnetization behavior at some temperatures. (c) Heat-capacity as a function of temperature in different fields. The lines through the data points serve as guides to the eyes.

Figure 3:
(a) Dielectric constant as a function of temperature below 50 K, in the absence of external magnetic field. Inset shows the curve for a wider temperature range. The curves in the presence of external magnetic fields are shown in (b). The arrow indicates the way the curves shift. In (c), the change in the dielectric constant as a function of magnetic field at selected temperatures are shown and the curves are very weakly hysteretic.

Figure 4:
(a) Temperature dependence of dc-bias current ($I_{bias}$), obtained as described in the text for different rates of heating, with an electric field of 3.5 kV/cm. (b) Remnant polarization is plotted in the presence of external fields. Insets in (a) $I_{bias}$ plot for –3.5 kV/cm and (b) the pyrocurrent behavior for $E$= 3.5kV/cm.

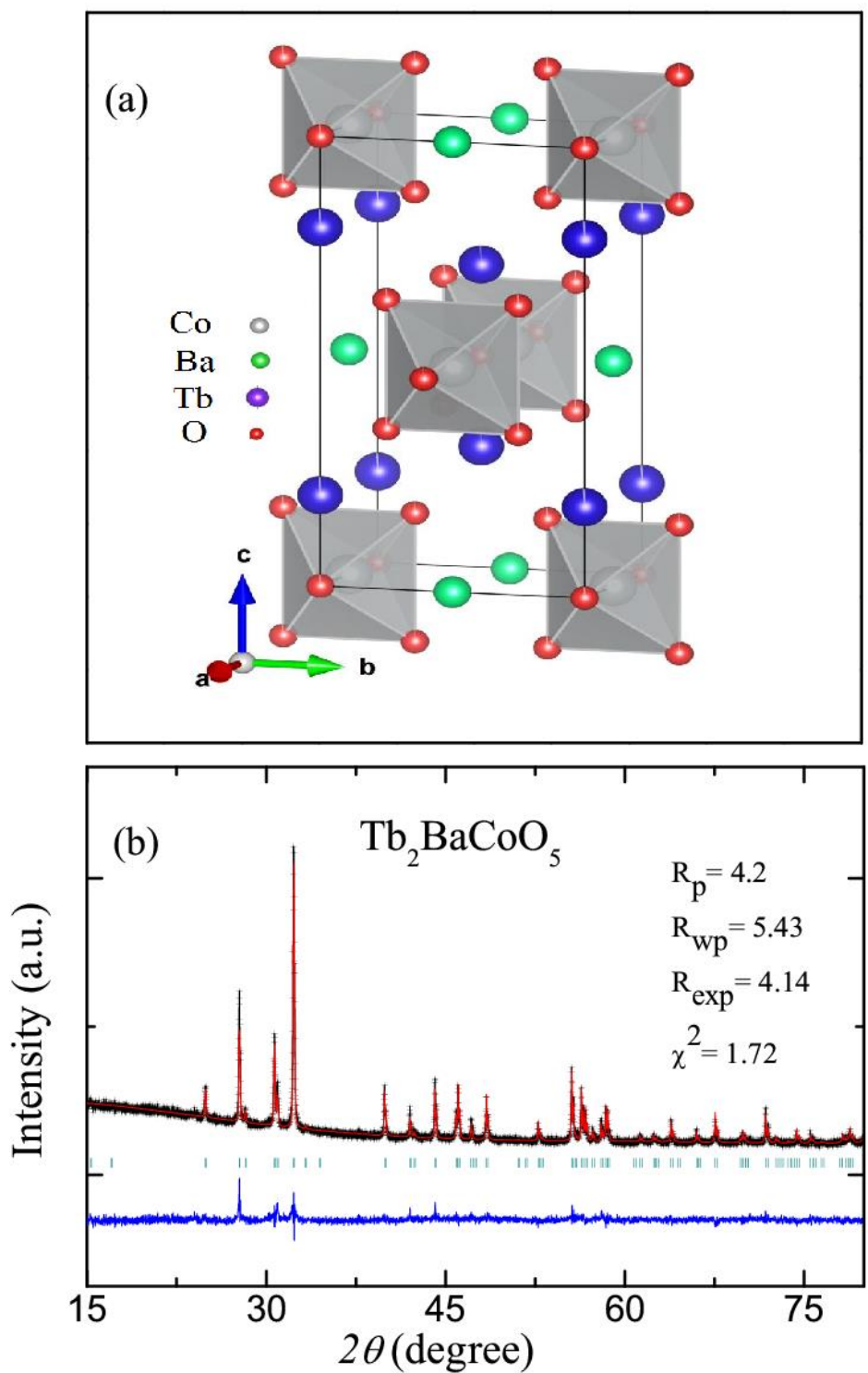

Figure 1

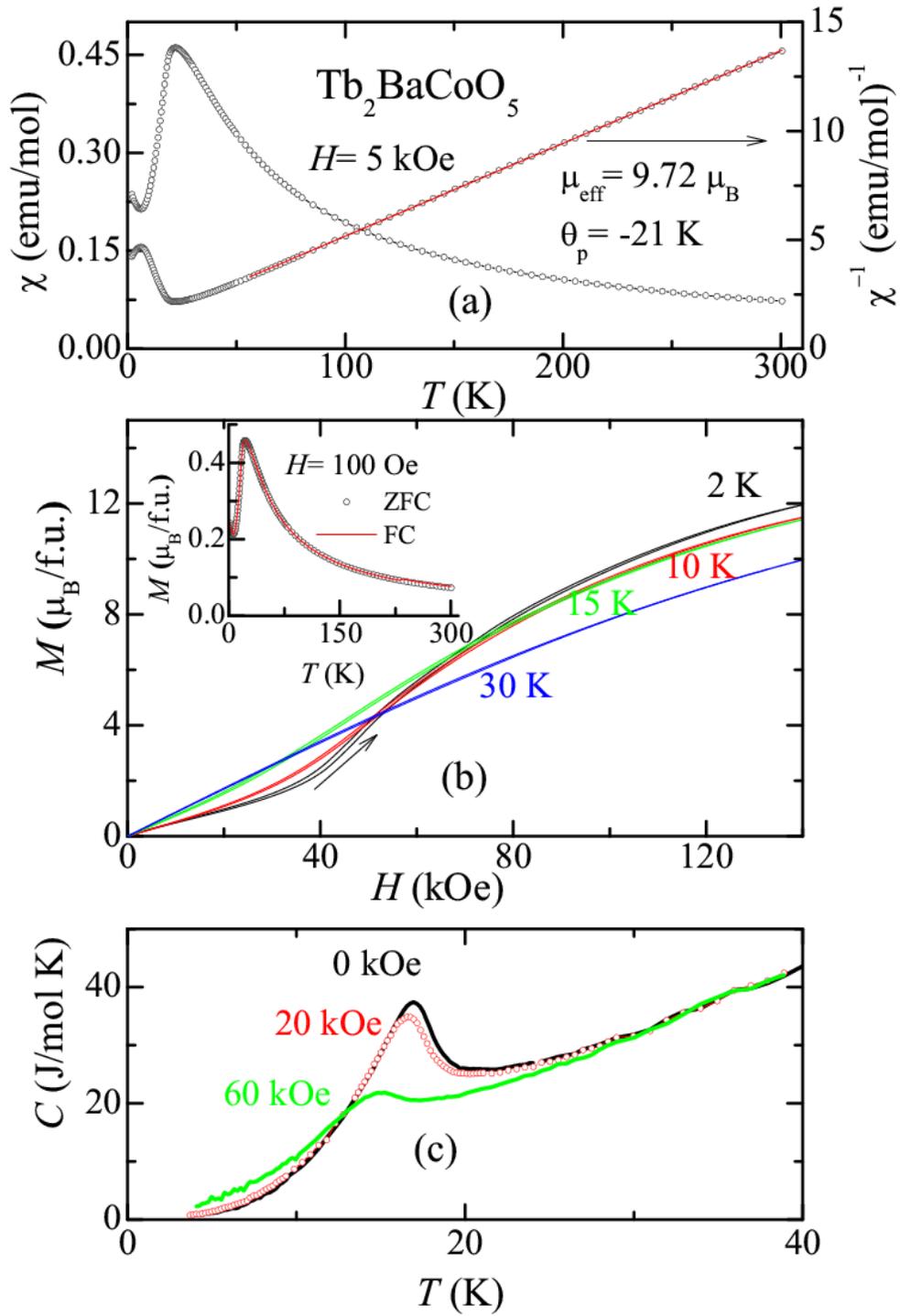

Figure 2

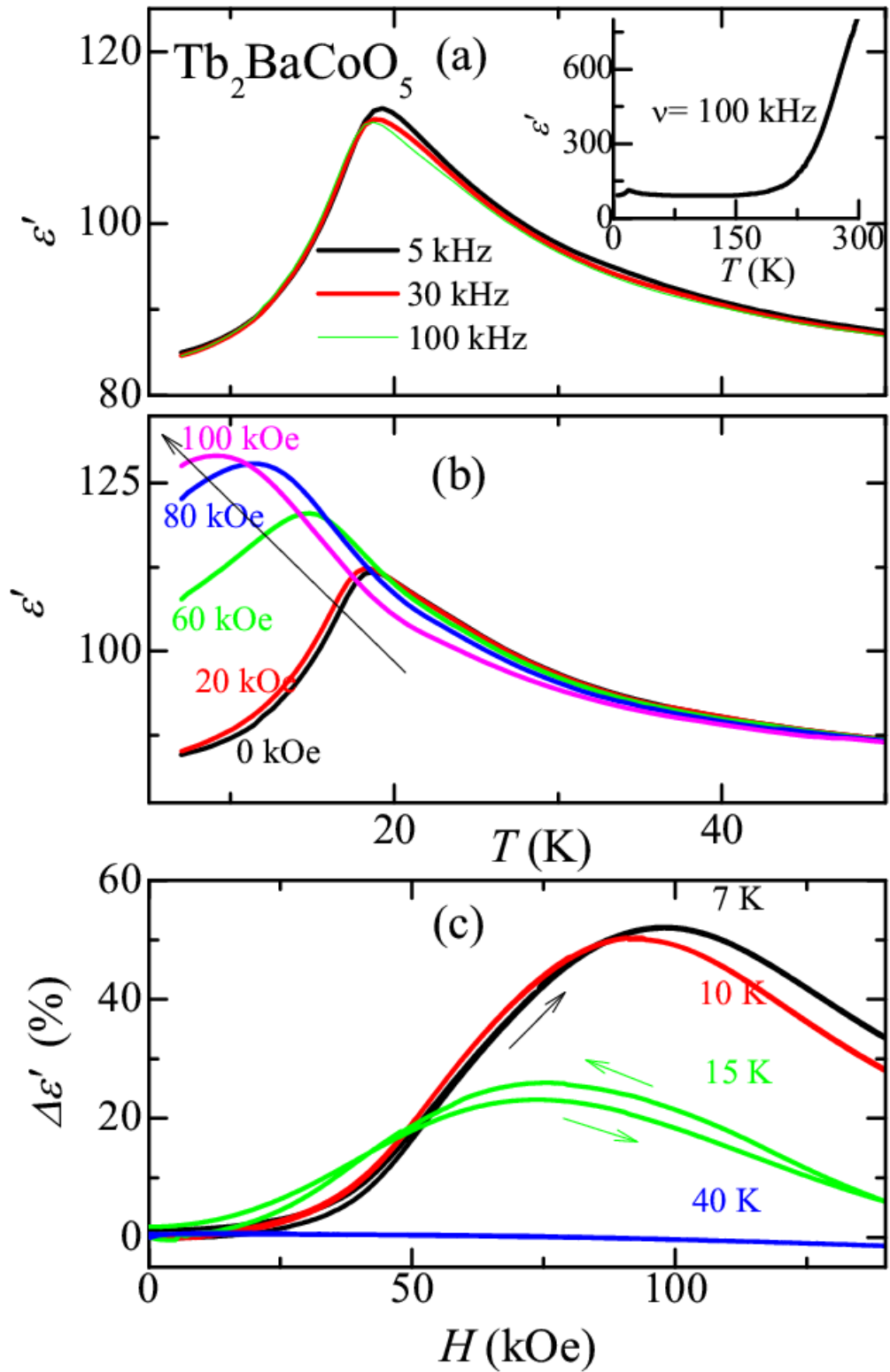

Figure 3

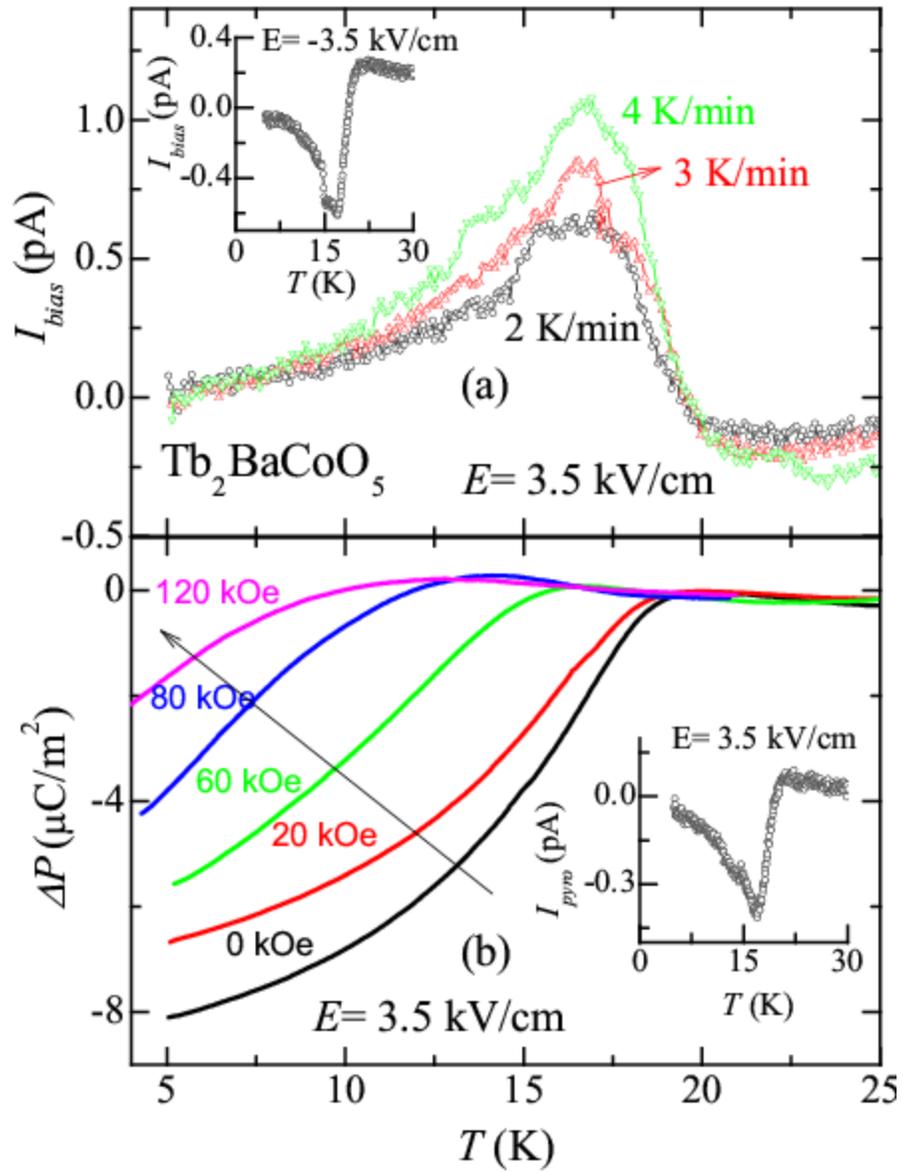

Figure 4